\documentclass[showpacs,prb,letterpaper,floatfix,nobalancelastpage,twocolumn]{revtex4}
\usepackage{dcolumn}% Align table columns on decimal point
\usepackage{bm}% bold math
\usepackage{amsmath}
\usepackage{bm,braket}
\usepackage{color}
\usepackage{float}
\usepackage{graphicx,subfigure}
\usepackage{amssymb}
\usepackage{sidecap}
\usepackage{float} % new added to located the figure at the given position
\begin{document}

% change equation skips
%\abovedisplayskip=8pt
%\abovedisplayshortskip=8pt
%\belowdisplayskip=8pt
%\belowdisplayshortskip=8pt
%\arraycolsep=100pt
%%%%%%%%%%%%%%%%%% title page information %%%%%%%%%%%%%%%%%%

% Title of the article
%\title{
%Resonance fluorescence of a two-level system coupled to a single metal nanoparticle: %Sampling a metallic--photon-reservoir  with strong  internal coupling effects}

%\title{
% Mollow triplet sampling of a metallic--photon-reservoir  with strong  internal %coupling}

\title{Accessing quantum nanoplasmonics in a hybrid quantum-dot  metal nanosystem: Mollow triplet of a quantum dot near a metal nanoparticle}

\author{Rong-Chun Ge$^1$,  C. Van Vlack$^1$, P. Yao$^2$,  Jeff. F. Young$^3$, and S. Hughes$^1$}
\email{shughes@physics.queensu.ca}
\affiliation{$^1$Department of Physics, Engineering Physics and Astronomy,
Queen's University, Kingston, Ontario, Canada K7L 3N6\\
$^2$Department of Optics and Optical Engineering,  University of Science and Technology of China, 230026, People's  Republic of China\\
$^3$Department of Physics and Astronomy, University of British Columbia, 6224 Agricultural Rd., Vancouver, B.C., V6T 1Z1, Canada
}
% \author{C. Van Vlack}
% \affiliation{Department of Physics, Engineering Physics and Astronomy,
% Queen's University, Kingston, Ontario, Canada K7L 3N6}
% \author{P. Yao}
% \affiliation{Department of Optics and Optical Engineering,
%University of Science and Technology of China, 230026, People�s
%Republic of China.}
% \author{Others?}

\begin{abstract}

We present a theoretical study of the resonance fluorescence spectra of an optically driven quantum dot placed near a single metal nanoparticle.  The metallic reservoir coupling is calculated for an 8-nm metal nanoparticle using a time-convolutionless master equation approach where the exact photon reservoir function is included using Green function theory. By exciting the system coherently near the nanoparticle dipole mode, we show that the driven Mollow spectrum becomes highly asymmetric due to internal coupling effects with higher-order plasmons. We also highlight  regimes of resonance squeezing and broadening as well as spectral reshaping through light propagation.
Our master equation technique can be applied
to any arbitrary material system, including lossy inhomogeneous structures, where mode expansion techniques  are known to break down.
% on the QD-cw laser detuning.
% which shows substantial differences from experimental results.
\end{abstract}

\pacs{42.50.Pq, 78.67.Bf, 73.20.Mf}
%42.50.Pq, 78.67.Bf, 73.20.Mf
\maketitle

%%%%%%%%%%%%%%%%%%%%%%%%%%  body  %%%%%%%%%%%%%%%%%%%%%%%%%%

\section{Introduction}
%Recently, there has been much interest in quantum plasmonics, where
%quantum light-matter interactions can be exploited in the vicinity
%of a metallic surface.
The study  of quantum light-matter interactions near metals
can  be used to explore fundamental quantum optical regimes
such as modified spontaneous emission~\cite{Novotny:PRL06}
and  the strong coupling regime~\cite{Hohenester:PRB08,Savasta:Nano10,Vanvlack:PRB12}, with
applications ranging from single-photon transistors~\cite{Sorensen:Nature07}
to plasmon lasing and spacing~\cite{Naginov:Nature09,Oulton:Nature09,Bergman:PRL03}. Metal structures enable surface plasmon polaritons which give rise to pronounced resonances in a similar way to high-Q (quality factor) cavity structures. However, metals are significantly more complicated to model because of material losses, and, e.g., standard mode expansion techniques that are well used in quantum optics theory typically fail. This motivates the need for quantum optics models that can be applied to metallic environments, and  can expect to access new excitation regimes that are unique to plasmonic systems.

Recently, there has  been interest in the coherent excitation of a single atom or quantum dot (QD) near a metal surface. As is well known from atomic optics, a resonantly driven atom (or QD) can yield a ``Mollow
triplet'' for the incoherent spectrum if the coherent Rabi oscillations have
a frequency that is larger than the decay rates in the system~\cite{Mollow,Carmichael:Book1}.
Ridolfo {\em et al.}~\cite{Savasta:PRL2010} modelled  QD metal-nanoparticle (MNP) interactions through  a   master equation (ME) approach by assuming a single Lorentzian response for the metal and estimated the dot-metal coupling parameters from electromagnetic simulations; such an approach is useful but  restricted since it essentially ignores the higher-order plasmon modes and cannot be used if the dot is too close to the metal surface---typically restricted to separation distances greater than a radius of the particle~\cite{Novotny:PRL06,Vanvlack:PRB12}
or else coupling to higher-order plasmons becomes important.
Gonzalez-Tudela {\em et al.}~\cite{Tudela:PRB2010} employed a time-convolutionless (i.e., time local) ME approach
 and explored the coupling for a driven QD near a metal planar surface; this latter method allows one to incorporate the full non-Lorentzian lineshape of the metal reservoir for QDs that are close to the surface, however the effects of internal coupling~\cite{carmichael} and 
spectral reshaping due to QD/MNP-to-detector propagation
were neglected.
Internal coupling refers to  coupling to the photon reservoir at the dressed-state resonances, since in general the scattering rates and radiative coupling depend on the driving field~\cite{Tanas:JMO:2001,carmichael}. For a suitable MNP environment, the range of energy shifts of the dressed states can be substantial compared to the range of energies over which the LDOS varies.
%Internal coupling effects have  been shown to be important for a driven QD coupled to bulk acoustic phonons~\cite{roy_hughes}, resulting
%in excitation induced damping.

\begin{figure}[b]
\centering\includegraphics[width=.99\columnwidth]{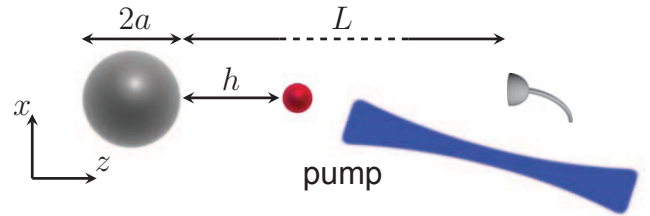}
\caption{(Color online)
Schematic showing a pump field exciting the QD metal-nanoparticle system (not to scale), resulting in resonance fluorescence that can be detected in the far field. The radius of the MNP is $a=8~$nm. The distance from the MNP to the detector is varied in the results presented below. }
%\vspace{-0.55cm}
\label{fig:fig1}
\end{figure}

%  and were been pointed out
%in earlier works by
%Carmichael and Walls~\cite{carmichael}, and
%Kowalewska-Kud\l ask and Tana\'s~\cite{Tanas:JMO:2001}.

In this work, we introduce a powerful ME technique that allows one to model the quantum light-matter interactions for any general photonic reservoir function, including lossy inhomogeneous structures. The only restriction we use is the second-order Born approximation, which is valid for the weak QD-plasmon coupling regime that 
we consider.
We apply this approach to study the incoherent spectrum that is detected when a QD is driven resonantly near an 8-nm MNP.
A schematic of this excitation scheme is shown  in Fig.~\ref{fig:fig1}. In the strong-field excitation regime, we compute the fluorescence spectrum at a detector remotely located  from the driven QD-MNP system, fully taking into account the effects of light propagation and optical quenching. We demonstrate that, as the field strength of the drive is increased, the ensuing Mollow triplets becomes highly asymmetric. The Mollow triplets gives direct access to the regime of quantum nanoplamonics and contains signatures of the MNP's photon bath function.
%We also show that this spectral asymmetry comes from the need to include %internal coupling effects.

The paper is organized as follows. In Sec.~\ref{theory}  we introduce the theory and ME technique for modelling a coherently driven QD in the vicinity of a MNP. We also present an expression for the incoherent spectrum in terms of the medium Green functions which are computed exactly.
In Sec.~\ref{results}, we present calculations of the Green functions and incoherent spectra for various pump intensities and pump laser detunings.
The ensuing Mollow spectrum is seen to be highly asymmetric and we show the
importance of including internal coupling effects. We also study the effect
of the QD-MNP separation and observe squeezing and anti-squeezing of the spectral resonances. Theoretical expressions for the linewidths in terms
of the LDOS help to explain the physics.
 Conclusions
are offered in Sec.~\ref{conclusions}. We also   include an Appendix that presents the optical Bloch equations for this system and discuss useful analytical limits of the incoherent spectrum.

\section{Theory}
\label{theory}
\subsection{Photon Green Function }
%Photonic Green function of the MNP
%{\em Theory.---}
We first introduce the classical photonic Green function  of the MNP, where the nanoparticle is assumed to be in air. Defining the MNP complex permittivity as
$\varepsilon_{\rm MNP}({\bf r},\omega)
=\varepsilon_R({\bf r},\omega) + i\varepsilon_I({\bf r},\omega)$, then
the  photon  Green function  satisfies the follow  equation ($\omega$ is implicit):
\begin{align}
\nabla\times\nabla\times{\bf G}({\bf r},{\bf r}')-k_{0}^{2} \varepsilon_{\rm MNP}({\bf r}) {\bf G}({\bf r},{\bf r}')=k_{0}^2\delta({\bf r}-{\bf r}'){\bf I},
\end{align}
where $k_0=\omega/c$, $\mu=1$, and ${\bf I}$ is the unit dyadic.
We use a Drude model for a silver MNP, where $\varepsilon(\omega)
= \varepsilon_\infty -\omega_m^2/(\omega^2-i\gamma_m\omega)$,
with $\varepsilon_\infty=6$, $\omega_m$=7.9~eV
and $\gamma_m=51~$meV. For a spherical MNP, the  Green function is computed exactly \cite{Li:IEEE1994}. The MNP Green  function can be understood in terms of having contributions from
the fundamental dipole-like plasmon mode and a reservoir
of higher-order plasmon modes. The dipole mode
propagates to the far field, but the higher-order modes only couple in the near field. We consider an  8-nm radius MNP, where a QD center is at some distance $h$  above the MNP surface.  We assume there is a detector, at position  ${\bf r}_D)$, that is 10~$\mu$m from the surface (i.e., in the far field). 

\begin{figure}[t]
\centering\includegraphics[width=.99\columnwidth]{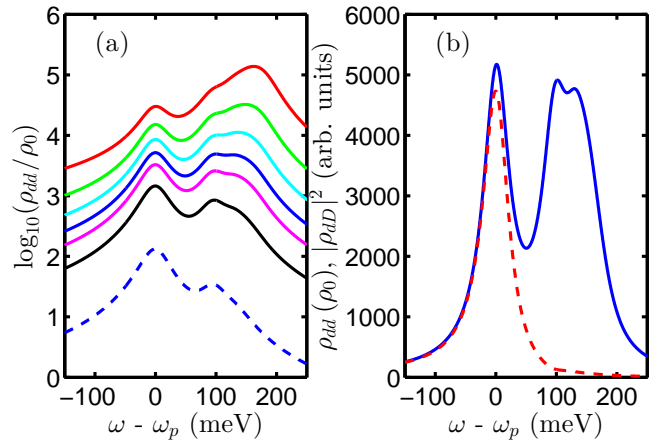}
\caption{(Color online) LDOS $\rho_{dd}$ at the point where the ${\bf r}_{\rm dot}$ located around a 8-nm silver MNP. (a) logarithm of LDOS at various locations: (top to bottom)  2~nm (red, upper), 3~nm (green), 4~nm (cyan), 5~nm (blue), 6~nm (magenta), 8~nm (black),  and 16-nm (blue dashed, lower) away from the surface of the MNP (from top to bottom), respectively. (b)  LDOS $\rho_{dd}$ (blue) at ${\bf r}_{\rm dot}$ situated 5 nm away from the surface of the MNP, and the $\rho_{dD} = |{\bf G}_{zz}({\bf r}_D,{\bf r}_d;\omega)|$ (red), where ${\bf r}_D$ (detector position) is 10$~\mu$m above the dot
position.
The fundamental plasmon resonance of the dipole mode is $\omega_p\approx 2.79~$eV. {\bf }}
\label{fig:fig2}
\end{figure}

In  Fig.~\ref{fig:fig2}(a),  we show examples of
the logarithm of $z-$projected  LDOS,
 $\rho_{dd}\equiv\rho_{zz} = {\rm Im}[{\bf G}_{zz}({\bf r}_d,{\bf r}_d;\omega)]$ in units of $\rho_{0}$, where $\rho_{0}=k_0^3/(6\pi\varepsilon_0)$ is the imaginary part of ${\bf G}_{ii}({\bf r},{\bf r};\omega)$ for free space.
For the smallest separations of 2~nm ($h/R = 0.25$), then a QD exciton
can be strongly coupled to the MNP resulting in vacuum Rabi splitting~\cite{Vanvlack:PRB12};
this strong coupling regime becomes accessible by coupling to the higher-order plasmon peaks rather than at the weaker dipole mode peak of the LDOS; 
one can see from Fig.~\ref{fig:fig2}(a) that the corresponding LDOS (red solid, upper)  at high order plasmon modes is several times larger than at the dipole mode peak. For larger spatial separations between the QD and MNP surface,
the LDOS decreased rapidly with the higher-order plasmon mode decreasing more rapidly than the dipole mode; the QD-MNP strong coupling regime is barely resolvable 
at $h=3~$nm, and is completely lost for separation distances
of more than $h=4~$nm or $h/R>0.5$.  For the majority of our calculations we will consider $h=5~$nm, though later we will also study the light-matter interactions at $h=8~$nm;
for the  spatial separation
of $h=5$~nm, we have verified that a second-order Born approximation is valid
and we will use this coupling regime to introduce a general master equation below.
From Fig.~\ref{fig:fig2}(a), one can see that even for $h=16~$nm ($h/R=2$), there is still  an influence from the higher-order plasmon modes and the use of a single Lorentzian model would fail in general.
In Fig.~\ref{fig:fig2}(b), we show the LDOS at the $h=5~$nm in a linear scale
and also  the magnitude of the non-local photon ``propagator,''
$|\rho_{dD}| = |{\bf G}_{zz}({\bf r}_D,{\bf r}_d;\omega)|$.
Clearly the reservoir function, $\rho_{dd}$, cannot be
described by a single Lorentzian lineshape; in contrast, the propagator, $\rho_{dD}$, is mainly influenced
by the dipole mode~\cite{Sun:APL10,Novotny:PRL06,Vanvlack:PRB12}
and is thus much closer to a single Lorenzian response.

\subsection{Master Equation}

For the QD  interactions, we consider a two-level system (artificial atom) in the dipole approximation,
interacting with a general lossy and inhomogeneous structure (the MNP). The total Hamiltonian of the coupled system can be written as~\cite{Welsch:PRA99,Suttorp:EPL04}
\begin{align}
H& = \hbar \int d{\bf r} \int_0^\infty d\omega  \, \omega \,{\bf f}^\dagger({\bf r},\omega) {\bf f}({\bf r},\omega) +\hbar\omega_{x}\sigma^+\sigma^-
\nonumber \\
&-\left [\sigma^+ \int_0^\infty \! d\omega\, {\bf d} \cdot {\bf E}({\bf r}_{d},\omega) + {\rm H.c.}\right ] +
%\hbar\eta_x(\sigma^+e^{-i\omega_L}+\sigma^-e^{i\omega_L}),
H_{\rm drive} ,
\end{align}
where $\sigma^+/\sigma^-$ are the Pauli operators of the exciton (electron-hole pair), $\omega_x$ is the resonance of the exciton,
${\bf d}$ is the dipole of the exciton~\cite{Notes1}, ${\bf f}/{\bf f}^\dagger$ are the boson field operators, and the rotating-wave approximation has been applied (i.e., the counter-rotating-wave term has been dropped). The  electric-field operator (not including the pump field) is defined through~\cite{Welsch:PRA99,Gruner,Dung}
\begin{align}
{\bf E}({\bf r},\omega) = \frac{1}{\varepsilon_0} \int d{\bf r}'\,
{\bf G}({\bf r},{\bf r}';\omega)\cdot\,
%${\bf G}({\bf r},{\bf r}',\omega)$, whose imaginary part at ${\bf r}={\bf %r}'$ is related to the LDOS.
\sqrt{\frac{\hbar\varepsilon_0}{\pi}  \varepsilon_I({\bf r}',\omega)}\,{\bf f}({\bf r}',\omega),
\end{align}
 and for convenience we have separated the pump Hamiltonian, defined through $H_{\rm drive}=\frac{\hbar\Omega}{2}\left(\sigma^+e^{-i\omega_L}+\sigma^-e^{i\omega_L}\right)$, with the effective Rabi field $\Omega=\braket{{\bf E}_{\rm pump}({\bf r}_d)}\cdot{\bf d}/\hbar$. The pump field contains the direct pumping term plus the (dominant) scattered field from the MNP, \begin{align}
&{\bf E}_{\rm pump}(
{\bf r}_d,\omega_L)={\bf E}_0({\bf r}_d,\omega_L)\,\,+\nonumber\\
&\,\,\,\,\,\int_{V_{\rm MNP}}d{\bf r}'{\bf G}({\bf r}_d,{\bf r}';\omega_L)\left[\varepsilon_{\rm MNP}(\omega_L)-1\right]{\bf E}_0({\bf r}',\omega_L),
\end{align}
where ${\bf E}_0({\bf r}_d,\omega_L)$ is the incident field operator and for a large driving field we can treat the Rabi field classically (i.e., as a ``c-number"). Note that the spatial integration is carried out over the volume of the MNP; in this way, we recognize a plasmonic enhancement factor of
$\eta_p\equiv1+\int_{V_{\rm MNP}}d{\rm r}'{\bf G}({\bf r}_d,{\bf r}';\omega_L)\left[\varepsilon_{\rm MNP}(\omega_L)-1\right]$,
 where ${\bf G}$ is the total Green function of the environment including the presence of the MNP. Exciting near the dipole resonance with a $z$-polarized incident field, we estimate that $\Omega\equiv\eta_p\Omega_0$ is at least one order of magnitude larger than $\Omega_0=\braket{{\bf E}_0({\bf r}_d)}\cdot{\bf d}/\hbar$. Thus the incident Rabi field is substantially enhanced by the MNP plasmonic response~\cite{Ratchford:Nano11,Waks:PRA10}.
%which satisfies the Kramers-Kronig relations and is valid for any arbitrary %medium.
%

In a frame rotating at the laser
frequency $\omega_L$, we rewrite the above Hamiltonian as
$H=H_S+H_R+H_{I}$,
where the system, reservoir (or bath), and system-reservoir interaction terms  are defined through
\begin{subequations}
\begin{align}
H_S& =  \hbar(\omega_{x}-\omega_L)\sigma^+\sigma^-
+ \hbar\eta_x(\sigma^++\sigma^-),
%-\left [i\sigma^+ \int_0^\infty d\omega\, {\bf E}({\bf r}_{d},\omega) + {\rm H.c.}\right ] +
%\hbar\eta_x(\sigma^+e^{-i\omega_L}+\sigma^-e^{i\omega_L}),
%H_{\rm drive},
 \\
H_{R}& = \hbar \int d{\bf r} \int_0^\infty d\omega  \, \omega
%(\omega-\omega_L)
 \,{\bf f}^\dagger({\bf r},\omega) {\bf f}({\bf r},\omega),
 \\
  H_{I}&  =  -\left [\sigma^+ e^{i\omega_L t} \int_0^\infty d\omega\,{\bf d} \cdot {\bf E}({\bf r}_{d},\omega) + {\rm H.c.}\right ] ,
\end{align}
\end{subequations}
and then transform the Hamiltonian  using
$\tilde H \rightarrow  U^\dagger(t) H U(t)$
with $U(t)=\exp[-i(H_{S}+H_R)t/\hbar]$, where the tilde denotes the interaction picture. We next manipulate the system-reservoir interactions to derive a ME with the reservoir interaction included within a Born approximation.
%For numerical convenience,
A common choice for the ME  is the time-convolutionless form~\cite{me_book_1}.
To second order in the interaction,  one has
\begin{align}
\label{ME_I}
\!\!\frac{\partial \tilde{\rho}(t)}{\partial t} =
-\frac{1}{\hbar^{2}}\int^{t}_{0}d\tau\,\rm{Tr}_{R}\left \{\left [\tilde{H}_{I}(t),\left [\tilde{H}_{I}(t-\tau), \tilde{\rho}(t)\rho_{R}\right ] \right ]\right \},
\end{align}
where $\tilde\rho$ is the reduced density operator in the interaction picture, and $\rho_R= \rho_{R}(0)$ is the density operator of the photonic reservoir  which is assumed to be initially in thermal
equilibrium.
Using the  bath approximation,
${\rm Tr}_R[
 {\bf f}_i({\bf r},\omega)
 {\bf f}_i^\dagger({\bf r}',\omega')
\rho_R]
= [\bar n(\omega)+1]
\delta({\bf r}-{\bf r}')\delta(\omega-\omega')
 $ and
 ${\rm Tr}_R[
 {\bf f}^\dagger_i({\bf r},\omega)
 {\bf f}_i({\bf r}',\omega')
\rho_R]
= \bar n(\omega)
\delta({\bf r}-{\bf r}')\delta(\omega-\omega')$,
we  consider the zero-temperature bath limit  (i.e., $\bar n(\omega)=0$), which is appropriate for optical frequencies.
Exploiting the relation  $\int d{\bf s}\, \varepsilon_I({\bf s},\omega)
{\bf G}({\bf r},{\bf s},\omega) {\bf G}^*({\bf s},{\bf r}',\omega) = {\rm Im}[{\bf G}({\bf r},{\bf r}',\omega)],$
%coming back
transforming back to the
Schr\"odinger picture, and carrying out the trace over the photon reservoir~\cite{Tanas:JMO:2001}, we derive the generalized ME:
 \begin{align}
 \label{eq:ME1}
  \frac{\partial {\rho}}{\partial t}    & =
  \frac{1}{i\hbar}[H_S,\rho] + \int_0^t d\tau   \left \{{\tilde J_{\rm ph}(\tau)}   [-\sigma^+ \sigma^-(-\tau)\rho\, +  \right . \nonumber \\
& \left .  \phantom{\tilde J_{\rm ph}} +\sigma^-(-\tau)\rho\sigma^+  ] + {\rm H.c.} \right \}
   +{\cal L}_{\rm pure}(\rho),
 \end{align}
where
%${\rm H.c}$ refers to Hermitian conjugate and
 $\tilde J_{\rm ph}(\tau) =  \int_0^\infty d\omega J_{\rm ph}(\omega)   e^{i(\omega_L-\omega)\tau}$, with
 the photon-reservoir spectral function given by
$J_{\rm ph}(\omega) \equiv \frac{{\bf d} \cdot {\rm Im}[{\bf G}({\bf r}_{ d},{\bf r}_{ d};\omega)] {\bf  \cdot d}}{\pi\hbar\varepsilon_0}$, and $ {\cal L}_{\rm pure}(\rho)  =
\frac{{\gamma}'}{2}(2{\sigma}_{11}\rho{\sigma}_{11}
-{\sigma}_{11}{\sigma}_{11}\rho-\rho{\sigma}_{11}{\sigma}_{11})$,
 where $\sigma_{11}=\sigma^+\sigma^-$ and $\gamma'$ is the exciton pure dephasing rate.
The time-dependent operators, which are defined through
 $\sigma^{\pm}(-\tau)=e^{-iH_S\tau/\hbar} \sigma^{\pm} e^{iH_S\tau/\hbar}$,
highlight that the scattering rates are pump-field dependent in general, since different dressed states can sample different parts of the photonic LDOS~\cite{Lewenstein:PRA88}.
%For QD structures it is also necessary to incorporate the effects of pure dephasing on the zero phonon line; the
%relevant phenomenological superoperator is defined through

For on-resonance driving (i.e., $\omega_L\,=\,\omega_x$), then
$\sigma^{\pm}(\tau)=\frac{\sigma^{\pm}(0)}{2}\left[1+\cos{\Omega\tau}\right]+\frac{\sigma{\mp}(0)}{2}
\left[1-\cos{\Omega\tau}\right]\pm\sigma^z(0)i\sin{\Omega\tau}$, which shows explicitly the formation of new bath-mediated scattering processes such as incoherent excitation and pure dephasing, in addition to modified radiative decay. For numerical calculations, Eq.~(\ref{eq:ME1}) captures non-Markovian dynamics, but for our MNP spectral functions, we find no evidence for non-Markovian behavior on the ensuing spectrum, so we can safely extend the upper time integration on Eq.~(\ref{eq:ME1}) to infinity. We subsequently obtain a useful analytic form for the ME,
\begin{align}
&\frac{\partial \rho}{\partial t}=\frac{1}{i\hbar}\left[H_S,\rho\right]+\frac{\Gamma(\Omega)}{2}\left(2\sigma^-\rho\sigma^+-
\sigma^+\sigma^-\rho-\rho\sigma^+\sigma^-\right)\nonumber\\
&+{\cal L}_{\rm pure}(\rho)+N(\Omega)\left[\sigma^+\sigma^-,\rho\right]+M(\Omega)\left[\sigma^+,\sigma^z\rho\right]\nonumber\\
&+M^*(\Omega)\left[\rho\sigma^z,\sigma^-\right]+K(\Omega)\sigma^+\rho\sigma^++K^*(\Omega)\sigma^-\rho\sigma^- \,,
\label{eq:ME2}
\end{align}
where the various parameters are defined as follows (see the Appendix for further details):
\begin{subequations}
\begin{align}
\Gamma(\Omega)& =\frac{{\bf d}\cdot {\rm Im}[{\bf G}(\omega_L-\Omega)+2{\bf G}(\omega_L)+{\bf G}(\omega_L+\Omega)]\cdot{\bf d}}{2\hbar\varepsilon_0},
\label{parametersa}
\\ N(\Omega)& =i\frac{{\bf d}\cdot {\rm Re}[{\bf G}(\omega_L-\Omega)+2{\bf G}(\omega_L)+{\bf G}(\omega_L+\Omega)]\cdot{\bf d}}{4\hbar\varepsilon_0},\\
 M(\Omega)& =i\frac{{\bf d}\cdot [{\bf G}(\omega_L-\Omega)-{\bf G}(\omega_L+\Omega)]\cdot{\bf d}}{4\hbar\varepsilon_0},\\
K(\Omega)& =i\frac{{\bf d}\cdot [{\bf G}(\omega_L-\Omega)-2{\bf G}(\omega_L)+{\bf G}(\omega_L+\Omega)]\cdot{\bf d}}{4\hbar\varepsilon_0} ,
\label{parametersd}
\end{align}
\end{subequations}
where ${\bf G}(\omega) \equiv {\bf G}({\bf r}_d,{\bf r}_d;\omega)$ and we  use the scattered part of the Green function~\cite{Notes2}.
Note that in the above equations we have included the principal value part~\cite{Tanas:JMO:2001} exactly. It is also interesting to note that similar terms appear for QDs that are coupled to an acoustic phonon bath~\cite{Ata:arXiv}.

%By neglecting internal coupling effects between the system Hamiltonian and the bath (e.g., replacing $\sigma^{\pm}(-\tau)$
%by $\sigma^{\pm}$), we recover previous TC ME approaches
%for modelling light-matter interactions near a metallic reservoir~\cite{Tudela:PRA2010}.
%Such an approach is valid only for weak driving, where one can then define the low-pump radiative decay rate through
%$
%\gamma(t) \approx
%2 \int_0^t  {\rm Re} [\tilde J_{\rm ph}'(\tau)]  d\tau,
%$
%where
%$\tilde J_{\rm ph}'(\tau) \approx  \int_0^\infty d\omega J_{\rm ph}(\omega)   e^{i(\omega_x-\omega)\tau}$,
%since for $\eta_x \ll \Delta_{xL}$,
%then $\sigma^{\pm}(-\tau) \approx e^{\mp i(\omega_x-\omega_L) \tau}\sigma^{\pm}$,
%and one can then replace replace $\sigma^{\pm}(-\tau)$ with $\sigma^{\pm}$ in Eq.~(\ref{eq:ME1}).

%Interestingly, we see that the LDOS at the exciton dictates the radiative %decay rate, even if the drive is off resonance (assuming that the drive %is weak).
%Using ${\bf G}$ for free space,
%we can obtain $\gamma_{0}$ and define the Purcell factor (modified spontaneous %emission rate factor),  as ${\rm PF}=\gamma/\gamma_{0}$. For $\omega_x=\omega_p$, % $\gamma(t)$ is shown in Fig.~\ref{fig:fig2}(b), which shows a rapid evolution %towards the Markovian average, reaching a substantial Purcell factor ($>5000$) %in less than 0.2~ps.

\subsection{Incoherent Spectrum }
%{\em Incoherent Spectrum.---}
%We finally  connect back to the general electric-field operator, as we have
%in mind to calculate the detected spectrum at some position ${\bf r}_D$ %away from the QD position.
To connect to experiments on resonance fluorescence, the detected spectrum will depend upon the position ${\bf r}_D$
as  highlighted above for the MNP
[via ${\bf G}({\bf r}_D,{\bf r}_{d})$].
% It is common to assume an emitted spectral function that is independent %of detector position,
%and to assume that the same spectral function couples locally (to the exciton) %and non-locally (via propagation to the detector)~\cite{Tanas:JMO:2001}. %For the case of the MNP,  one can reasonably assume that the only out-propagating %mode is from the Lorentzian-like dipole mode~\cite{Savasta:PRL2010,cole:MNP2011}.
%But for small dot-MNP separations, the LDOS is significantly different than %a single Lorentzian.
The spectrum is defined from $S({\bf r}_D,\omega)=
\braket{({\bf E}^{\rm scatt}({\bf r}_D,\omega))^\dagger{\bf E}^{\rm scatt}({\bf r}_D,\omega)}$, where the scattering field due to the presence of QD is~\cite{yao:PRB09}
\begin{align}
{\bf E}^{\rm scatt}({\bf r}_D,\omega)&=\frac{1}{\varepsilon_0} {\bf G}({\bf r}_D,{\bf r}_{ d},\omega) \cdot {\bf d}\,[\sigma^-(\omega) +\sigma^+(\omega)]. %\nonumber \\ &+ {\bf E}_h({\bf r}_D,\omega),
\end{align}
%with ${\bf E}_h$ the homogeneous solution of the field in the absence of %any dot; we will neglect this term as it has no contribution to the incoherent
%spectrum (i.e., the MNP is {\em bosonic}).
For continuous wave excitation,
it is common to define the incoherent spectrum as follows:
%and again working in the rotating frame,   the following  incoherent spectrum
%is{\bf }
\begin{align}
S_{0}(\omega)&\equiv
\lim_{t\rightarrow\infty}{\rm Re}\left [\int_0^\infty d\tau \left(\langle{\sigma}^+(t+\tau){\sigma}^-(t)\rangle - \right. \right.\nonumber\\
& \left. \left.\langle{\sigma}^{+}(t)\rangle\langle{\sigma}^{-}(t)
\rangle \right) e^{i(\omega_{L}-\omega) \tau}  \right ],
\label{eq:S1a}
\end{align}
%\begin{align}
%S_{0}(\omega)&\equiv
%\lim_{t\rightarrow\infty}{\rm Re}\left [\int_0^\infty d\tau  \right. \nonumber\\
%&  \left . \left  (\langle{\sigma}^+(t+\tau){\sigma}^-(t)\rangle -  \langle{\sigma}^{+}(t)\rangle\langle{\sigma}^{-}(t)
%\rangle \right ) e^{i(\omega_{L}-\omega) \tau}  \right ],
%\label{eq:S1a}
%\end{align}
where the latter term in the above equation subtracts  the
elastic (coherent) scattering from the pump field.
Unfortunately, though commonly used in quantum optics, this expression is not {\em valid}---especially for metals---it neglects spectral filtering effects via light propagation.  Including propagation and quenching in a self-consistent way, we  derive the following {\em detectable} spectrum,
\begin{align}
S_{p}({\bf r}_D,\omega)&=\frac{2}{\varepsilon_0}\left |{\bf d}\cdot{\bf G}({\bf r}_D,{\bf r}_{d},\omega)\right |^2 S_0(\omega) ,
\label{eq:S1b}
\end{align}
which highlights the essential role of the propagator.
%We stress that the former form, $S_0$, is far from general and is quite %limited.
%For example, one may wish to probe the positional dependence of the emission
%in the near to far field, where one must account for the
%non-Lorentzian coupling to the reservoir.
%Equations (\ref{eq:S1a})-(\ref{eq:S1b}), however, can fully account non-Lorentzian %decay. In fact,
%It can be applied to compute the spectrum at any spatial position.
Equations (\ref{eq:S1a})-(\ref{eq:S1b}), together with (\ref{eq:ME1}), constitute our main results
with which to investigate resonance fluorescence of the exciton in the
vicinity of a MNP. Importantly, the MNP reservoir function is included exactly and the field operators are properly quantized. An explicit expression for the analytical spectrum is given in the Appendix, from which we obtain the full-width at half-maximum (FWHM) of the Mollow triplet center and sideband resonance, respectively:
\begin{align}
\label{eq:Rate1}
\Gamma_{\rm center}&(\Omega)  \approx \frac{\gamma'}{2}+\frac{\pi}{2}\left[J_{\rm ph}(\omega_L-\Omega)+J_{\rm ph}(\omega_L + \Omega)\right],
\\
\label{eq:Rate2}
\Gamma_{\rm side}&(\Omega)  =  \frac{\gamma'}{4} + \nonumber\\ &\frac{\pi}{4}\left[J_{\rm ph}(\omega_L-\Omega)+4J_{\rm ph}(\omega_L)+J_{\rm ph}(\omega_L+\Omega)\right].
\end{align}
We  recognize that the center line width is only affected by the projected LDOS (since $J_{\rm ph}(\omega) \propto {\bf d}\cdot{\rm Im}\left[{\bf G}({\bf r}_d,{\bf r}_d;\omega)\right]\cdot{\bf d}$) at the Mollow sidebands, while the sideband width depends on a linear combination at all three dressed resonance; while such effects have been predicted before for atomic and dielectric system~\cite{Florescu:PRA04,Keitel:OC95}, the effects are usually small and to the best of our knowledge have never been measured nor predicted for a lossy metal environment.

%Equation (\ref{eq:S1b}) can be considered the real-space and lossy-media generalization
%of the spectrum used in  Ref.~\onlinecite{Tanas:JMO:2001}.
% We also
% stress that is is common to make a Markov approximation on
% the out propagation so that the spectrum would follow $|{\bf G}({\bf r}_D,{\bf r}_a,\omega_0)|^2$. We make no Markov approximation in deriving Eq.~(\ref{eq:S1}).

\begin{figure}[t]
\centering\includegraphics[width=.99\columnwidth]{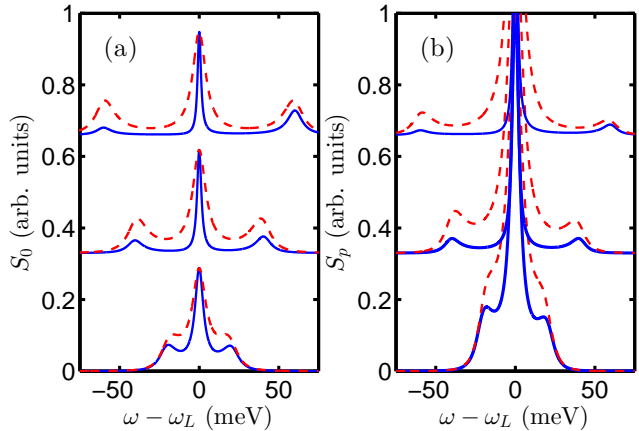}
\caption{(Color online)
Calculated spectra for increasing Rabi fields (bottom to top: $\Omega= 20,40,60~$meV), with $\omega_L=\omega_x=\omega_p$. (a) Spectra without propagation effects, $S_0$, with (blue solid) and without internal coupling effects (red dashed). (b) Corresponding spectra with propagation effects, $S_p$, which is the spectra that would be observed in the far field (via the plasmon dipole mode);  to better see the sidebands, we have zoomed into the lower part of the spectra.
}
\label{fig:fig3}
\end{figure}

\section{Results}
\label{results}
\subsection{Asymmetric  Mollow triplets}

%{\em Calculations of the Mollow Triplets.---}
To compute the resonance fluorescence spectra, we assume a QD dipole moment of $d\equiv |d|=30~$Debye and consider a pump field that excites the exciton with an effective pump rate, $\Omega$. Importantly, we have the dot in a spatial position where it necessarily feels the influence of the high-order plasmon modes, and consequently, we will show that such a system is an excellent environment in which to study generalized reservoir coupling. We then solve the steady state density matrix and use the quantum regression theorem (or formula~\cite{me_book_1}) to obtain the two-time correlation function and thus the spectrum.

In Fig.~\ref{fig:fig3}, we display the pump-dependent Mollow spectra for the example case of $\omega_L\,=\omega_x=\omega_p$,  using $\gamma'=0.1~$meV ~\cite{Notes3}.
 To better clarify the role of
internal coupling effects, we  show two sets of
results, with and without the time dependence of the $\sigma^{\pm}(-\tau)$ terms
in Eq.~(\ref{eq:ME1}) [i.e., we set ${\bf G}(\omega_L\pm\Omega)={\bf G}(\omega_L)$ in Eq.~(\ref{eq:ME2})]. In Fig.~\ref{fig:fig3}(a), we plot
the bare spectrum with (blue solid line) and without (red dashed line) time dependence, and recognize that there is  no asymmetry when the internal coupling term is turned off; although there can be a small lamb shift, we find that this is a negligible effect. These findings are consistent with previous results~\cite{Savasta:PRL2010}. However, with internal coupling, we observe a clear asymmetry in the Mollow triplets for sufficiently large driving fields; this can be explained by the complex energy (or frequency) dependence of the LDOS.  For each sideband of the triplet, there is a one-to-one correspondence between the height of the sideband and the corresponding total transition probability between the dressed states,  which is proportional to the LDOS at that  energy [see Eqs.~(\ref{eq:Rate1})-(\ref{eq:Rate2})], at the location of the QD.  The large asymmetry in the LDOS shown in Fig.~\ref{fig:fig2} therefore
 manifests itself in the 
 asymmetric strength of the Mollow triplets for sufficiently large drive fields. For the case without internal coupling, assuming large drives, then  only the LDOS at the frequency of the pump field matters and the corresponding Mollow triplet is symmetric. We also observe significant narrowing of the resonances, consistent with Eqs.~(\ref{eq:Rate1})-(\ref{eq:Rate2})]. In Fig.~\ref{fig:fig3}(b),
we observe similar features for $S_p(\omega)$, but now the spectra are further
reshaped due to photon propagation and quenching effects; specifically, without propagation effects the MTs are symmetric without internal coupling, and asymmetric with IC; with propagation effects, both are asymmetric, but in opposite senses.

\begin{figure}[t]
\centering\includegraphics[width=.99\columnwidth]{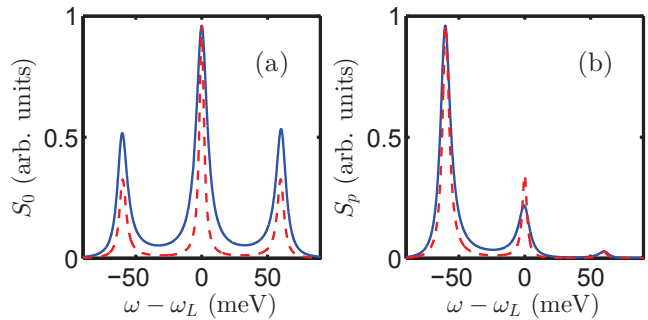}
\caption{(Color online) As in Fig.~\ref{fig:fig3}, using $\Omega=60~$meV, but now the pump frequency is $60$~meV above the fundamental plasmon mode: $\omega_L=\omega_p+60$~meV. (a) Spectra without propagation effects, $S_0$, with (blue solid) and without internal coupling effects (red dashed). (b) Corresponding spectra with propagation effects. Here the fundamental plasmon mode is seen to relatively enhance the lower Mollow sideband.}
\label{fig:fig4}
\end{figure}

Next, we investigate the case when the pump field is resonant with the exciton, but off resonant with the dipole mode of the MNP, and show that the spectral filtering effect can be used to selectively enhance the features of the Mollow sidebands.  In Fig.~\ref{fig:fig4}, we show the high-field solution when the pump field is now $60$~meV blueshifted with respect to the plasmon dipole mode ($\omega_L=\omega_x=\omega_p+60~$meV). Here the photon propagator has changed the spectrum of the scattering field dramatically. Due to the fact that the left sideband predominates over both other peaks, it should be more accessible in an experiment. Figure~\ref{fig:fig4}(a) shows the following features:  ($i$) the induced-asymmetry from internal coupling is now minor  since the LDOS at $\omega_L\pm60~$meV is similar, and ($ii$) there is  significant spectral broadening when internal coupling is included; this broadening effect is anticipated from our theory as the Mollow sideband resonances now sample a larger LDOS [see Fig.~\ref{fig:fig1} and Eqs.~(\ref{eq:Rate1})-(\ref{eq:Rate2})]. We stress that such effects are not possible in a Lorentzian-decay medium (e.g., a single mode cavity).

%In addition, we include pure dephasing.

\subsection{Position dependence of the Mollow triplet}

Due to the fact that high-order plasmons are strongly confined near the surface of the MNP, one may expect that as the distance between the QD and MNP is increased, then a simple single Lorentzian may be valid. In addition, it is useful to know by how much the Mollow triplet features change if one moves
the spatial position of the QDs by a few nm.
As is shown in Fig.~\ref{fig:fig2}(a), even if the distance between the QD and the surface of the MNP is as large as the radius of MNP, the higher-order plasmon modes still have a significant impact and the failure of a single Lorenzian model is to be expected. 
In fact, we find the photonic resevoir of QD induced by the MNP will always display some non-Loretzian characteristics.

\begin{figure}[t]
\centering\includegraphics[width=.99\columnwidth]{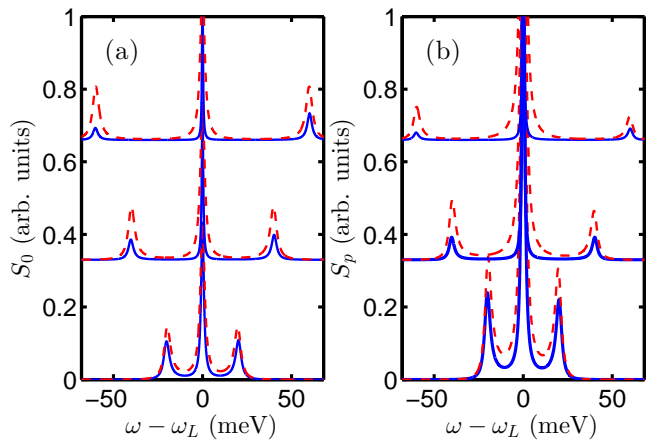}
\caption{(Color online)
Calculated spectra for the QD located 8-nm away from the surface of the MNP with increasing Rabi fields (bottom to top: $\Omega= 20,40,60~$meV), and $\omega_L=\omega_x=\omega_p$. (a) Spectra without propagation effects, $S_0$, with (blue solid) and without internal coupling effects (red dashed). (b) Corresponding spectra with propagation effects, $S_p$.
}
\label{fig:fig6}
\end{figure}

Figure \ref{fig:fig6} presents the Mollow spectra with the pump frequency $\omega_L=\omega_x=\omega_p$. Figure \ref{fig:fig6}(a) shows the bare spectrum with (blue, solid line) and without (red, dashed line) internal coupling, and clearly there is still an asymmetry between the Mollow sidebands; however, since the LDOS values are reduced the magnitude of the sidebands are suppressed and the radiative decay rates are much smaller (less broadening). In Fig.~\ref{fig:fig6}(b), we show the detectable spectrum in the far field, which again shows significant reshaping of the spectrum  due to photon propagation and quenching effects.
However,  as the spatial distance between the QD and MNP is further increased, the contribution of higher-order plasmon modes becomes smaller and smaller; eventually we find that when the QD is placed 16~nm  ($2R$) away from the MNP surface  [blue dashed in Fig.~\ref{fig:fig2}(a)], then the magnitude of the LDOS can be reasonably described by the dipole mode only as the LDOS contribution from the  higher-order plasmon modes is about one order of magnitude smaller than the dipole mode peak LDOS; in this larger separation regime, the MNP can be effectively described by the dipole model as also discussed elsewhere~\cite{Vanvlack:PRB12,Jacob:APL12}.

%{\bf Due to the fact, high order plasmon strongly confined to the surface of MNP, as the distance between the QD and the MNP %become larger the LDOS of the photon reservoir at high energy decrease much fast than those at the low energy; As a result, %the width of both the central and sidebands decrease, and the degree of asymmetry becomes smaller. Even $h$ is as large as %the radius of the MNP ($8~$nm), the metal reservoir still could not be described by single Lorentzain.}

\section{Conclusions}
\label{conclusions}
%{\em Summary.---}
We have introduced a general ME approach for modelling quantum light-matter
interactions for a driven atom or QD in the vicinity of a metallic nanoparticle. The exact reservoir function and propagator are obtained from Green function theory and used directly
in the ME formalism. We used this approach to model the Mollow spectrum
as a function of drive strength, and demonstrated that clear spectral asymmetries and non-trivial linewidth variations
can be seen for suitably large drives.
We also investigated several different pump excitation frequencies and QD positions and find rich coupling behaviour. While our master equation formalism is useful for exploring regimes of quantum nanoplasmonics, the  techniques  are  general and can, assuming the validity of the second-order Born approximation, model the spectrum from a driven QD in any general photonic environment, including hybrid metal-photonic-crystal systems~\cite{barth}
and metamaterials~\cite{yao:PRB09,Jacob:APL12}.\\

\section{Acknowledgments}
This work was supported by the National Sciences and Engineering Research Council of Canada,
the Canadian Institute for Advanced Research,   the
National Key Basic Research Program of China under grant
No.2012CB922003, and the National Natural Science Foundation of China under grant No.61177053.

%%%%%%%\vfill\eject
%\columnbreak

%{Derivation of the fluorescence spectrum and Mollow triplet linewidths}

\begin{widetext}
\vspace{-0.2cm}
\appendix*
\section{Derivation of the Incoherent Spectrum and Full-Width at Half Maximum of the Mollow Triplet Resonances}
Using the master equation, Eq.~(\ref{eq:ME2}), we derive the following Bloch equations
%%%%\lipsum[1]
%%%%\begin{widetext}
\begin{align}
\frac{d \langle\sigma^{+}\rangle}{d t}=&-\left (\frac{\Gamma(\Omega)+\gamma'}{2}+N(\Omega)\right)\langle\sigma^{+}\rangle+
K^*(\Omega)\langle\sigma^{-}\rangle
- i\frac{\Omega}{2}\langle\sigma^z\rangle-M^*(\Omega),\\
\frac{d \langle\sigma^{-}\rangle}{d t}=&-\left (\frac{\Gamma(\Omega)+\gamma'}{2}-N(\Omega)\right)\langle\sigma^{-}\rangle+
K(\Omega)\langle\sigma^{+}\rangle + i\frac{\Omega}{2}\langle\sigma^z\rangle-M(\Omega),\\
\frac{d \langle\sigma^z\rangle}{d t}=&-\big(2M(\Omega)+i\Omega\big)\langle\sigma^+\rangle -\big(2M^*(\Omega)
-i\Omega\big)\langle\sigma^-\rangle -\Gamma(\Omega)(\langle\sigma^z\rangle+1),
\end{align}
%%%%\end{widetext}
%%%%\lipsum[1]
with parameters
$\Gamma(\Omega)$ defined through Eq.~(\ref{parametersa})-(\ref{parametersd}) in the main text of the manuscript.
These Bloch equations can be solved exactly, which is the approach we have used in the main text above, with no approximation. However, it is useful to look at certain limit of the corresponding solution for the spectrum, $S_0(\omega)$. This allows us to make a clearer connection to the underlying physics of the resulting spectral linewidths. Neglecting the terms from the real part of the Green function (since these only cause spectral shifts through the Lamb- and Stark- shifts), the steady-state solutions of the Bloch equations are
as follows:\begin{align}
\langle\sigma^+\rangle_{\rm ss}=&
i\frac{\Omega\left(\frac{\Gamma(\Omega)+\gamma'}{2}-K_r(\Omega)-4\frac{M_r^2(\Omega)}
{\Gamma(\Omega)}\right)}
{2\left (\frac{\Gamma(\Omega)+\gamma'}{2}-K_r(\Omega)\right)\left(\frac{\Gamma(\Omega)+\gamma'}{2}+
K_r(\Omega)+
\frac{\Omega^2}{\Gamma(\Omega)}\right)}-\frac{M_r(\Omega)}{\frac{\Gamma(\Omega)+\gamma'}{2}-K_r(\Omega)
},\\
\langle\sigma^-\rangle_{\rm ss}=&\langle\sigma^+\rangle^*_{\rm ss},\\
\langle\sigma^z\rangle_{\rm ss}=&-4\Gamma^{-1}(\Omega){\rm Re}\left[\left(M_r(\Omega)+i\frac{\Omega}{2}\right)\langle\sigma^+\rangle_{\rm ss}\right]-1,
\end{align}
where $K_r(\Omega)={\rm Re}[K(\Omega)]$ and $M_r(\Omega)={\rm Re}[M(\Omega)]$.
The incoherent spectrum, without propagation effects (see main text), is given by
%\begin{widetext}
\begin{align}
S_{0}(\omega)&\equiv
\lim_{t\rightarrow\infty}{\rm Re}\left [\int_0^\infty d\tau \left(\langle{\sigma}^+(t+\tau){\sigma}^-(t)\rangle -  \langle{\sigma}^{+}(t)\rangle\langle{\sigma}^{-}(t)
\rangle\right) e^{i(\omega_{L}-\omega) \tau}  \right ]\nonumber\\
&={\rm Re}\left\{\frac{d(0)\times\frac{-2K_r(\Omega)\delta\omega-2M_r(\Omega)\Omega-
i[\Omega^2+2K_r(\Omega)\Gamma(\Omega)]}{-i\delta\omega+\frac{\Gamma(\Omega)+\gamma'}{2}
-K_r(\Omega)
}+\Omega h(0)-
2[\delta\omega+i\Gamma(\Omega)]f(0)}{-2\delta\omega\left(K_r(\Omega)+\frac{3\Gamma(\Omega)+\gamma'}
{2}\right)+2i\left[-\Omega^2+\delta\omega^2-\left(K_r(\Omega)+\frac{\Gamma(\Omega)+\gamma'}{2}\right)
\Gamma(\Omega)\right]}\right\},
\end{align}
%\end{widetext}
with $d(0)=f(0)+g(0)$, $f(0)\equiv\langle\delta\sigma^+\delta\sigma^-\rangle$, $g(0)\equiv\langle\delta\sigma^+\delta\sigma^+\rangle$, and $h(0)\equiv\langle\delta\sigma^+\delta\sigma^z\rangle,$ respectively. These terms are obtained from the steady-state Bloch-equation solutions, via
\begin{align}
f(0)=&\frac12(1+\langle\sigma^z\rangle_{ss}-2\langle\sigma^-\rangle_{\rm ss}\langle\sigma^+\rangle_{\rm ss}), \ \ \
g(0)=-\langle\sigma^+\rangle_{\rm ss}^2, \ \ \
h(0)=-\langle\sigma^+\rangle_{\rm ss}(1+\langle\sigma^z\rangle_{\rm ss}).\nonumber
\end{align}
In the strong pump limit (i.e., $\Omega\gg \gamma', \, |\Gamma(\Omega)|, \, |N(\Omega)|, \, |M(\Omega)|, \, |K(\Omega)|$),  the full-width at half maximum (FWHM) values of the Mollow triplet resonance are obtained from the imaginary parts of
\begin{align}
&\left[\frac{\Gamma(\Omega)+\gamma'}{2}-i\delta\omega
-K_r(\Omega)\right]\Bigg\{2i\left[\delta\omega^2-\Omega^2-\bigg(K_r(\Omega)+ \right.
%\nonumber\\
&\left.\left.\frac{\Gamma(\Omega)+\gamma'}{2}\right)
\Gamma(\Omega)\right]-2\delta\omega\left[K_r(\Omega)+\frac{3\Gamma(\Omega)+\gamma'}
{2}\right]\Bigg\}=0.
\end{align}
The corresponding roots are easily obtained:
\begin{align}
\delta\omega_0=&-i\left(\frac{\Gamma(\Omega)+\gamma'}{2}
-K_r(\Omega)\right),\\
\delta\omega_{\pm}=&\pm\frac{\sqrt{4\left[\Omega^2+\Gamma(\Omega)B(\Omega)\right]
-\left[\Gamma(\Omega)+B(\Omega)\right]^2}}{2} -i\left[\Gamma(\Omega)+B(\Omega)\right],
\end{align}
with
$B(\Omega)=\frac{\Gamma(\Omega)+\gamma'}{2}+K_r(\Omega)$.
In the strong field limit, the real parts of the roots correspond to $\omega_L$,$\omega_L\pm\Omega$, at these Mollow triplet  resonance, and the FWHM of the spectral linewidths are
\begin{align}
&\Gamma_{\rm center}(\Omega)  = \frac{\gamma'}{2}+\pi
\frac{J_{\rm ph}(\omega_L-\Omega)+J_{\rm ph}(\omega_L + \Omega)}{2},\\
&\Gamma_{\rm side}(\Omega)  = \pi\frac{J_{\rm ph}(\omega_L-\Omega)+4J_{\rm ph}(\omega_L)+J_{\rm ph}(\omega_L+\Omega)}{4}+ \frac{\gamma'}{4} ,
\end{align}
which explicitly show the role of the three LDOS values at the dressed-state resonance.

%\appendix

%\subsection{Dressed State Approach}
It is also useful to compare the above bare-state approach with an approximate stressed-state approach in the
secular approximation, which has been used before in the context of coupling to generalized reservoirs~\cite{Keitel:OC95,Florescu:PRA04}. The Mollow triplet can then be explained from the energy level scheme in the dressed-state picture, where the Mollow central peak is due to the evolution of $\langle\sigma^z_{\rm dress}(t)\rangle$, and the sidebands are related to the relaxation of the  dipole operators $\langle\sigma^{\pm}_{\rm dress}(t)\rangle$. The dressed-state operators are related to the bare state operator through the following transformations,
%\begin{align}
$\sigma^{\pm}_{\rm dress}=\frac{1}{2}(\sigma^z+\sigma^{\mp}-\sigma^{\pm})$ and
$\sigma^{z}_{\rm dress}=\sigma^+ + \sigma^-$.
%\end{align}
Using bare state operators, and adopting the secular approximation~\cite{Florescu:PRA04}, we obtain
the following Bloch equations,\onecolumngrid
\begin{align}
\frac{d \langle\sigma^{\pm}\rangle}{d t}=&\mp i\frac{\Omega}{2}\langle\sigma^z\rangle-\pi\langle\sigma^{\pm}\rangle\frac{J_{\rm ph}(\omega_L-\Omega)+
2J_{\rm ph}(\omega_L)+J_{\rm ph}(\omega_L+\Omega)}{4}-\nonumber\\
&\pi\frac{J_{\rm ph}(\omega_L-\Omega)-2J_{\rm ph}(\omega_L)+J_{\rm ph}(\omega_L+\Omega)}{4}\langle\sigma^{\mp}\rangle+
\pi\frac{J_{\rm ph}(\omega_L-\Omega)-J_{\rm ph}(\omega_L+\Omega)}{4}-\frac{\gamma'}{2}\langle\sigma^{\pm}\rangle,\\
\frac{d \langle\sigma^z\rangle}{d t}=&\big(\pi\frac{J_{\rm ph}(\omega_L-\Omega)-J_{\rm ph}(\omega_L+\Omega)}{2}-i\Omega\big)\langle\sigma^+\rangle
+\big(\pi\frac{J_{\rm ph}(\omega_L-\Omega)-J_{\rm ph}(\omega_L+\Omega)}{2}
+i\Omega\big)\langle\sigma^-\rangle-\nonumber\\
&\pi\frac{J_{\rm ph}(\omega_L-\Omega)+2J_{\rm ph}(\omega_L)+J_{\rm ph}(\omega_L+\Omega)}{2}(\langle\sigma^z\rangle+1).
\end{align}
Thus the time evolution  of the average values for the dressed state operators are given by
\begin{align}
\frac{d \langle\sigma^{z}_{\rm dress}\rangle}{d t}=&-\langle\sigma^z_{\rm dress}\rangle\frac{\pi\left[J_{\rm ph}(\omega_L-\Omega)+J_{\rm ph}(\omega_L+\Omega)\right]+\gamma'}{2}+
\pi\frac{J_{\rm ph}(\omega_L-\Omega)-J_{\rm ph}(\omega_L+\Omega)}{2}, \\
\frac{d \langle\sigma^{\pm}_{\rm dress}\rangle}{d t}=&\, \langle \sigma^{\pm}_{\rm dress} \rangle\left(\pm i\Omega-\frac{\pi\left[J_{\rm ph}(\omega_L-\Omega)+4J_{\rm ph}(\omega_L)+J_{\rm ph}(\omega_L+\Omega)\right]+\gamma'}{4}\right)
-\nonumber\\
&\langle\sigma^{\mp}_{\rm dress}\rangle
\frac{\pi\left[J_{\rm ph}(\omega_L-\Omega)+J_{\rm ph}(\omega_L+\Omega)\right]+\gamma'}{4}+\langle\sigma^z_{\rm dress}\rangle
\pi\frac{J_{\rm ph}(\omega_L-\Omega)-J_{\rm ph}(\omega_L+\Omega)}{4}\nonumber\\
&-\pi\frac{J_{\rm ph}(\omega_L-\Omega)+2J_{\rm ph}(\omega_L)+J_{\rm ph}(\omega_L+\Omega)}{4},
\end{align}
from which we obtain the spectral linewidths,
\begin{align}
&\Gamma_{\rm pop}(\Omega) \equiv \Gamma_{\rm center}(\Omega)  = \frac{\gamma'}{2}+\pi
\frac{J_{\rm ph}(\omega_L-\Omega)+J_{\rm ph}(\omega_L + \Omega)}{2},
\\
&\Gamma_{\rm coh}(\Omega) \equiv \Gamma_{\rm side}(\Omega)  = \frac{\gamma'}{4} + \pi\frac{J_{\rm ph}(\omega_L-\Omega)+4J_{\rm ph}(\omega_L)+J_{\rm ph}(\omega_L+\Omega)}{4}.
\end{align}
As expected, these are in agreement with the more exact bare state approach when analyzed in the Mollow limit (which is similar to making the secular approximation). However, the advantage of Eq.~(\ref{eq:ME2}) is that it can be applied for all values of the pump field, so the secular approximation is not needed.
%\twocolumngrid
\end{widetext}

%%\lipsum
%%\begin{figure*}
%%\hrule  % if we include this command the will be a line just above this equation
%\vskip-1.9in
%%\end{widetext}
%%\lipsum

%%%%%%%%%%%%%%%%%%%%%%% References %%%%%%%%%%%%%%%%%%%%%%%%%


\begin{thebibliography}{[1]}
\bibitem{Notes2} Otherwise the Green function is divergence and this contribution only constributes to the vacuum Lamb shift which has been absorbed into the definition of $\omega_x$.
\bibitem{Notes3} Note that the numerical results  are essentially identical for  a larger rate of $\gamma'=1~$meV, since the plasmon coupling completely dominates the decay. Even for $\gamma'=10~$meV, the main features remain.
\end{thebibliography}
\end{document}